\documentclass[sigconf]{acmart}
\AtBeginDocument{%
  \providecommand\BibTeX{{%
    \normalfont B\kern-0.5em{\scshape i\kern-0.25em b}\kern-0.8em\TeX}}}

\setcopyright{acmcopyright}\acmConference[ICSEW'20]{IEEE/ACM 42nd International Conference on Software Engineering Workshops }{May 23--29, 2020}{Seoul, Republic of Korea}
\acmBooktitle{IEEE/ACM 42nd International Conference on Software Engineering Workshops (ICSEW'20), May 23--29, 2020, Seoul, Republic of Korea}

\usepackage{todonotes}
\usepackage{subfigure}
\usepackage{verbatim}
\begin{document}

\title{Improving The Effectiveness of Automatically Generated Test Suites Using Metamorphic Testing}

\author{Prashanta~Saha}
\email{prashantasaha@montana.edu}
\affiliation{%
  \institution{School of Computing, Montana State University}
   \city{Bozeman}
 \state{Montana}
 \country{USA}}
\author{Upulee~Kanewala}
\email{upulee.kanewala@montana.edu}
\affiliation{%
  \institution{School of Computing, Montana State University}
   \city{Bozeman}
 \state{Montana}
 \country{USA}
}


\begin{abstract}
Automated test generation has helped to reduce the cost of software testing. However, developing effective test oracles for these automatically generated test inputs is a challenging task. Therefore, most automated test generation tools use trivial oracles that reduce the fault detection effectiveness of these automatically generated test cases. In this work, we provide results of an empirical study showing that utilizing metamorphic relations can increase the fault detection effectiveness of automatically generated test cases.   
\end{abstract}




\maketitle

\section{Introduction}
\label{sec:intro}

Software testing is a costly activity yet essential to detect faults. Typically in testing, an \emph{oracle} is used to check whether the output produced for a given test input is correct or not ~\cite{10.1093/comjnl/25.4.465}. Much work has been done on automated test case generation, including the development of publicly available tools~\cite{Fraser:2011:EAT:2025113.2025179}. The main focus of this work has been on developing efficient methods to generate test inputs to achieve a particular target such as coverage and weak mutation score~\cite{Saha:2018:FDE:3193977.3193982}. However, there has been relatively less attention paid on utilizing effective oracles to improve the fault detection effectiveness of these automatically generated test cases.

\begin{figure*}[t]
  \begin{subfigure}[]{}
    \includegraphics[width=0.3\textwidth]{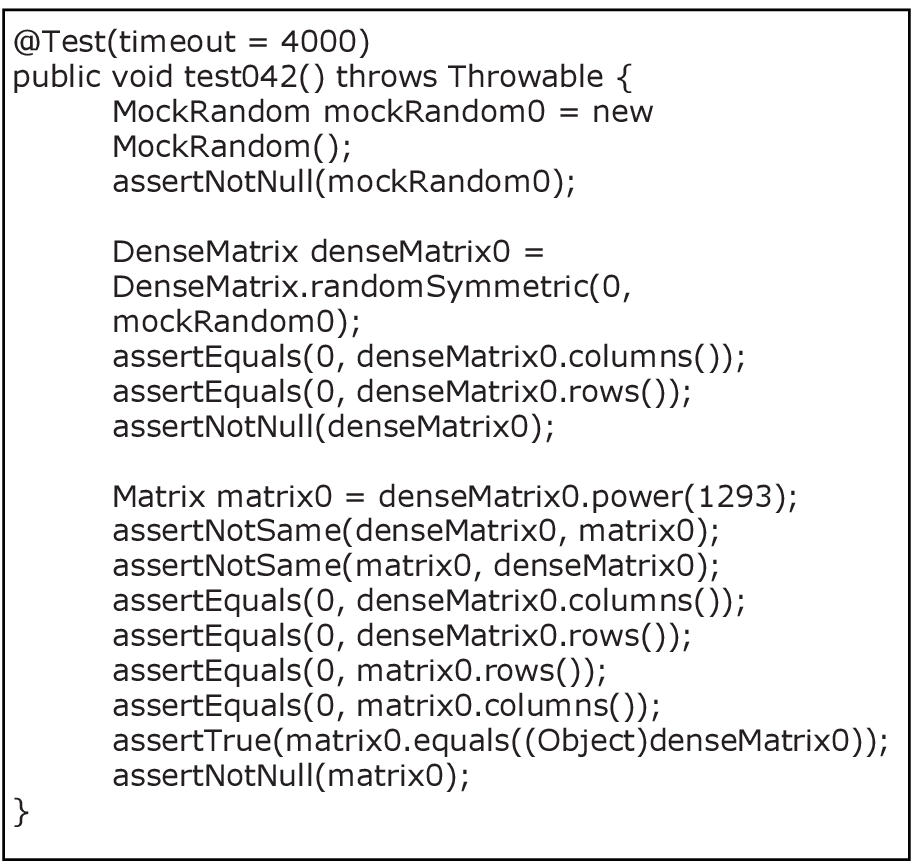}
    \label{fig:6a}
  \end{subfigure}
~
  \begin{subfigure}[]{}
    \includegraphics[width=0.3\textwidth]{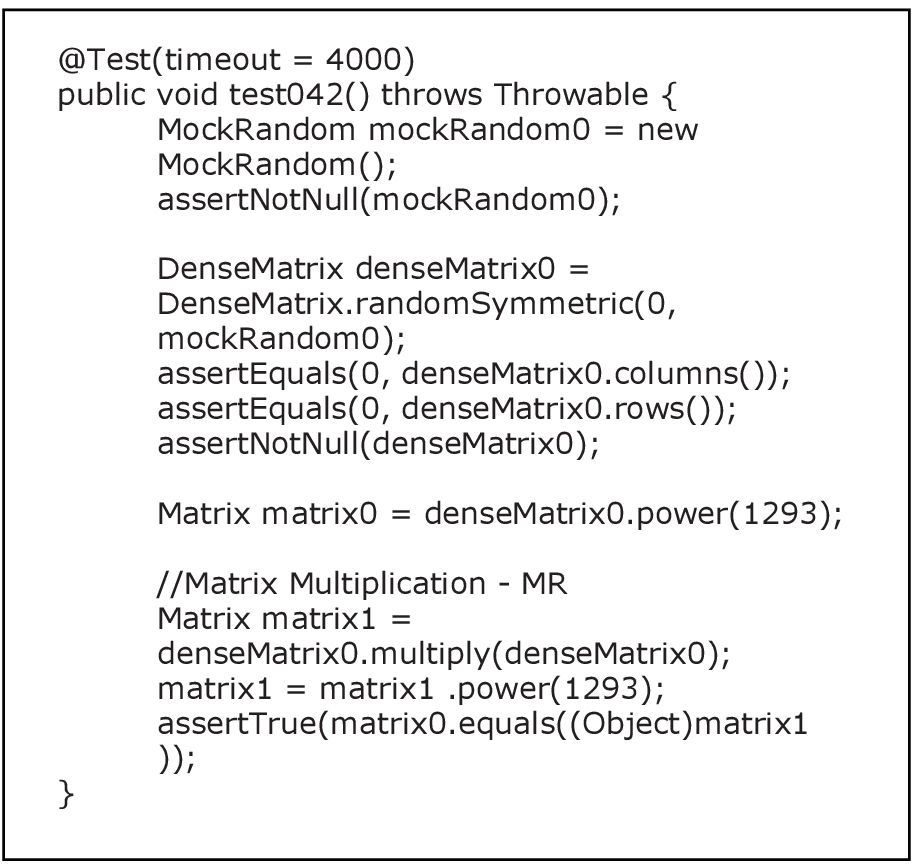}
    \label{fig:6b}
  \end{subfigure}
    \caption{(a) EvoSuite Generated Test Case , (b) Modified Test Case with MR in MT}
    \label{fig:6}
\end{figure*}

Metamorphic Testing (MT) is a technique proposed to alleviate
the oracle problem of software under test (SUT)~\cite{Chen1998MetamorphicTA}. This is based on the idea that most of the time it is easier to predict relations
between outputs of a program, than understanding its input-output
behavior. Such a relation is called a Metamorphic Relation (MR) in MT, and is a necessary property of the SUT that specifies
a relationship between multiple inputs and their outputs~\cite{Chen:2018:MTR:3177787.3143561}.

Automatically generated test suites have certain advantages over manually written test cases, in particular, saving human labor and time. Some work has shown that it is more effective to use test cases that are generated based on some coverage criteria rather than randomly generated test cases~\cite{Pacheco:2007:RFR:1297846.1297902}. However, due to the automated generation of test inputs, defining the oracles for these test inputs is a hard problem and faces the oracle problem. Thus, many of the automatically generated test cases would contain trivial oracles, such as the assert statements that we discussed above. This reduces the fault detection effectiveness of these test cases. Therefore, in this work, we investigate whether we can utilize MRs to improve the fault detection effectiveness of automatically generated test cases. For example, figure~\ref{fig:6}a is an EvoSuite generated test case for \textit{Power} method. This method powers a matrix of the given component (i.e. \textit{int} n) and returns the powered matrix. Though this test case has a code coverage of 100\% but the generated assert statements are weak to detect critical faults in the method. Because of the presence of such trivial oracles, the fault detection effectiveness of this test case is reduced. With Multiplication MR, we modified the current test case from figure~\ref{fig:6}b. We multiplied the source test case matrix with the same matrix. We ran the test case for the \textit{Power} method. Then we expected the resultant matrix from these two test cases are equal, or the follow-up output is higher than source output and compared them using assertion statements.

In this paper, we present the initial results of an empirical study conducted to evaluate the effectiveness of utilizing MRs with automatically generated test inputs. Our preliminary results show that MRs can help to increase the effectiveness of automatically generated test suites.
\section{Empirical Study}
\label{sec:tcg}
In this experiment we used 4 classes (\textit{Matrix.java, LeastSquaresSolver.java, ForwardBackSubstitutionSolver.java} and \textit{SquareRootSolver-
.java}) from \emph{la4j\footnote{\url{http://la4j.org/}}} (version 0.6.0) open-source Java library. la4j is a linear algebra library that provides matrix and vector implementations and algorithms and was one of the software packages used for evaluating the performance of automated testing tools. For each of these 4 classes, we used EvoSuite \cite{Fraser:2011:EAT:2025113.2025179} tool to generate test cases targeting line, branch, and weak mutation coverage. We have identified 16 MRs for the above 4 classes. These MRs are created based on common matrix operations (e.g., Transpose Matrix, Identity Matrix). We manually verified those input-output relationships of MRs with some sample values. Then we ran those MR modified source test cases (follow-up test cases) with automated source test inputs on the original programs and verified the MR properties again. If any MR did not hold for any test input, we excluded that MR for that particular input. 

We used mutation testing, in particular, PIT\footnote{https://pitest.org/} tool to  generate mutants, to measure the fault detection effectiveness of the test cases enhanced with MRs. We considered a mutant as "killed" when the MR violates the output relation and as "alive" when the relationship holds. We collected all the killed/alive information and calculated the mutation score and fault detection ratio for automated test suites and MRs.

\section{Preliminary Results and Conclusions}
\label{sec:results}

\begin{figure}[ht]
  \includegraphics[width=0.51\textwidth]{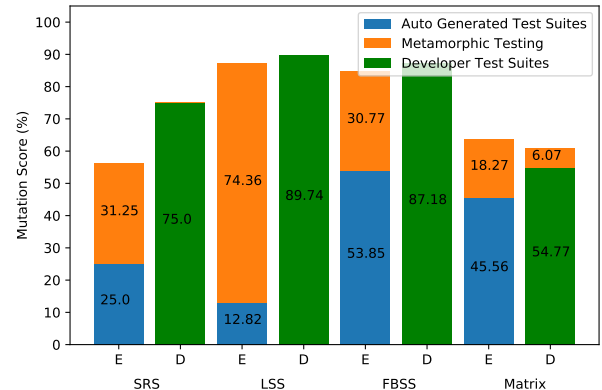}
  \caption{4 classes with Mutation score of automatically generated test suites and developer test suites,  and increase of mutation score with Metamorphic Testing. (SRS = SquareRootSolver, LSS = LeastSquaresSolver, FBSS = ForwardBackSubstitutionSolver, E = EvoSuite, D = Developer)}
  \label{fig:2}
\end{figure}

Figure \ref{fig:2} shows the fault detection effectiveness of EvoSuite generated test cases (orange), and the Evosuite test cases utilizing MRs (blue). We also show the fault detection effectiveness of developer written test cases. As shown in the results, there is a significant increase in the mutation score of when MRs are utilized with the automatically generated test suite. For two classes, the increase of the mutation score is 100\% higher than the automatically generated test suite. This preliminary result suggests that utilizing MRs with automatically generated test cases would improve the fault detection effectiveness. But for the case of  the developer test suite, there is no additional mutant killed by the MRs except for Matrix. This needs to be investigated further.

Our preliminary results are promising, and it suggests that MT can effectively improve the fault detection capability of automatically generated test suites. But we need a large scale implementation to  prove this claim further and to validate the correlation. We also need to find out the individual performance of MRs compared to the automatically generated test suites. 

\section{Acknowledgments}

This work is supported by award number 1656877 from the National Science Foundation. Any Opinions, findings and conclusions or recommendations expressed in this material are those of the author(s) and do not necessarily reflect those of the National Science Foundation.

\bibliographystyle{ACM-Reference-Format}
\bibliography{sample-base}

\appendix

\end{document}